\documentclass[a4paper,UKenglish,cleveref, autoref, thm-restate]{lipics-v2021}
\usepackage{cleveref}
\pdfoutput=1
\hideLIPIcs

\usepackage{xcolor}
\definecolor{keywordcolor}{rgb}{0, 0.1, 0.6}   
\definecolor{commentcolor}{rgb}{0.4, 0.4, 0.4}   
\definecolor{symbolcolor}{rgb}{0, 0.0, 0.0}    
\definecolor{tacticcolor}{rgb}{0, 0.3, 0.1}    
\definecolor{sortcolor}{rgb}{0, 0, 0}      
\definecolor{errorcolor}{rgb}{1, 0, 0}           
\definecolor{stringcolor}{rgb}{0.5, 0.3, 0.2}    

\usepackage{upgreek}

\def\lean{\lstinline[language=lean]}

\newcommand\ml[2]{\href{https://github.com/leanprover-community/mathlib4/blob/v4.17.0/Mathlib/Combinatorics/SimpleGraph/#1}{#2 \scriptsize{\faExternalLink*}}}
\newcommand\mlb[2]{\href{https://github.com/leanprover-community/mathlib4/blob/0d2016d6b2de4c164766a24bce95ca948950844c/Mathlib/Combinatorics/SimpleGraph/#1}{#2 {\scriptsize\faExternalLink*}}}

\title{Tutte's theorem as an educational formalization project}

\author{Pim {Otte}}{Utrecht University, Utrecht, The Netherlands \and \url{http://pim.otte.dev} }{p.j.otte@uu.nl}{https://orcid.org/0009-0001-4076-0616}{}

\authorrunning{P. Otte} 

\Copyright{Pim Otte}



\category{}

\relatedversion{}

\acknowledgements{I thank all mathlib contributors and maintainers and in particular 
Ya\"el Dillies, Bhavik Mehta and Kyle Miller for their useful comments on the formalization. 
I also thank my supervisors, Johan Commelin, Paige Randall North and Jim Portegies for their feedback and guidance
on the writing.}

\nolinenumbers

\EventEditors{John Q. Open and Joan R. Access}
\EventNoEds{2}
\EventLongTitle{42nd Conference on Very Important Topics (CVIT 2016)}
\EventShortTitle{CVIT 2016}
\EventAcronym{CVIT}
\EventYear{2016}
\EventDate{December 24--27, 2016}
\EventLocation{Little Whinging, United Kingdom}
\EventLogo{}
\SeriesVolume{42}
\ArticleNo{23}

\begin{document}

\maketitle

\begin{abstract}
  In this work, we present two results: The first result is the formalization of Tutte's theorem in Lean, a key
  theorem concerning matchings in graph theory. As this formalization is ready to be integrated in Lean's mathlib,
  it provides a valuable step in the path towards formalizing research-level mathematics in this area.
  The second result is a framework for doing educational formalization projects.
  This framework provides a structure to learn to formalize mathematics with minimal teacher input.
  This framework applies to both traditional academic settings and independent community-driven environments. 
  We demonstrate the framework's use by connecting it to the process of formalizing Tutte's theorem. 
\end{abstract}

\section{Introduction}\label{sec:intro}

In this work, we present the formalization of Tutte's theorem in Lean along with a framework for educational formalization projects.
Tutte's theorem characterizes the existence of perfect matchings in graphs. This theorem is a staple in undergraduate-level courses on
graph theory and is specifically core to the field of matching theory.
The formalization of this theorem and its proof techniques is an essential step toward the formalization of modern graph theory.
We present the formalization of Tutte's theorem as a polished formalized proof that has been nearly completely
integrated to Lean's mathlib.

With the advent of formalization in mathematics, the issue of training people to do formalization arises.
Since formalization is a optional technology in most of mathematics, there is an additional group of students
besides the traditional group of mathematics students: trained mathematicians who want to learn only formalization. 
Furthermore, there currently is an imbalance:
The group of capable teachers is limited compared to the group of potential students.
This raises the question: how do we teach both aspiring and current
mathematicians how to formalize a piece of mathematics, while minimizing teacher effort?

The first step in this training process is already facilitated by the various formalization communities
through the availability of several tutorials and entry-level materials. For example, in the Lean community,
some options include: Theorem Proving in Lean~\cite{avigad2021theorem}, 
Functional Programming in Lean~\cite{christiansen2022functional}
and Mathematics in Lean~\cite{avigad2020mathematics}.
This is why in this work we focus on the second step: Training beginners to be able to execute a larger formalization project.

As this second step, we present a framework for educational formalization projects. 
This framework consists of two phases: an initial formalization phase and a subsequent polishing phase.
We propose some necessary conditions under which this framework is appropriate, such as
the expected initial level of the student, requirements on teacher ability and requirements on project choice.
Using the formalization of Tutte's theorem as an example, we show how this framework worked in practice.
In this example, the author was the student and the Lean community functioned as a teacher.
This shows the flexibility of the framework: 
The teacher and student roles can also be assumed by volunteer mentors and learning enthusiasts;
therefore, throughout this work we use the terms `educational', `teacher' and `student' 
in their broadest possible senses: concerning the process of learning, anyone taking the role of teacher and anyone
taking the role of student, respectively.

In the remainder of this introduction, we discuss Tutte's theorem, its relevance for formalization,
summarize the framework, discuss related work and provide a reading guide for the rest of the paper.

\subsection{Tutte's theorem}

We provide the statement and a proof sketch of Tutte's theorem to give context for
 the formalization in \cref{sec:mathlib}.
In 1947, Tutte proved his characterization of finite graphs with perfect matchings \cite{tutte1947factorization}.
A perfect matching in a graph $G$ is a subgraph of $G$ such that all vertices have degree one, or equivalently
a partition of the vertices into pairs for which each pair is adjacent in $G$. 

\begin{theorem}[Tutte, 1947]
    A graph $G$ has a perfect matching if and only if for any subset $U \subset V$ the graph $G - U$ has at most
    $|U|$ components of odd size.
\end{theorem}

A subset $U \subseteq V$ such that the graph $G - U$ has more odd components than $|U|$ is referred to as a \emph{Tutte violator}. 
The necessity of the condition is fairly immediate: A Tutte violator immediately blocks the existence of a perfect matching, 
because at least one vertex in each odd component must be uniquely matched to a node in $U$. By the pigeonhole principle, this cannot occur.

Lov\'asz proved the sufficiency with the following structure (see Theorem 2.2.1 of Graph Theory by Diestel~\cite{diestel2017graph} for a full proof).
Argue by contraposition: Take a graph without a perfect matching and show a Tutte violator exists.
Without loss of generality, we can assume that adding any edge to this graph results in existence of a perfect matching.
In case the total number of vertices is odd, the empty set yields a Tutte violator.
Otherwise, we argue by contradiction and assume that the set of all vertices that are connected to all other vertices
(also referred to as the set of universal vertices) is not a Tutte violator.
We consider the graph obtained by removing these vertices and examine whether the remainder consists solely of cliques.
If it does, we explicitly construct a perfect matching by leveraging these cliques.
If it does not, then we obtain two perfect matchings on slightly bigger graphs and combine them to obtain a perfect matching on the original graph.
Combining these matchings involves both the symmetric difference of graphs, and cycles with the property that
exactly every other edge is also in a particular matching.
We treat the proof in more detail along with the formalization in \cref{sec:mathlib}.

Tutte's theorem is a worthwhile target for formalization for two reasons.
First, it is a core theorem in the area of matching theory. It is the precursor of the
Gallai--Edmonds Structure Theorem (see Theorem 3.2.1 of Matching Theory by Lov\'asz and Plummer~\cite{lovasz2009matching}), which yields the Gallai-Edmonds decomposition.
This decomposition basically pinpoints a canonical Tutte violator and is
a powerful tool that characterizes the structure of maximum matchings.
Second, combinatorics has a history of proofs assisted by brute force computer search,
which generally are considered to be contentious.
The proof of the Four Colour theorem by Appel and Haken~\cite{appel1977solution} serves as a prime example. 
The computer-checked proof in the Rocq prover\footnote{Formely known as Coq} of the Four Colour theorem in 2008 by Gonthier~\cite{gonthier2008formal} provided additional certainty
about the truth of this theorem.
Gonthier's version provided a formal proof of a sufficient brute force search along with the 
formal version of the more traditional mathematical part.
The proof of the boolean Pythagorean triples problem by Heule, Kullman and Marek~\cite{heule2016solving} uses a SAT solver 
and provides a certificate of unsatisfiability. This is much less contentious than a generic brute force search, because
the certificate can be checked.
Initial inspiration for choosing matching theory as an area for formalization stemmed from work by Otte~\cite{otte2014matchings},
where a proof of existence of exponentially many perfect matchings in cubic graphs (by Esperet, Kardo\v{s} and Kr\'al~\cite{esperet2012superlinear}) 
was extended using brute force search. While we do not consider this last work relevant enough to warrant full formalization,
these instances show that combinatorics as a whole and specifically graph theory are amenable to 
proofs using computer assistance, and therefore warrant a kind of formalization.

\subsection{The framework}

Formalization of mathematics is beginning to play a greater role in research-level mathematics.
Recent work by Gowers, Green, Manners and Tao~\cite{gowers2023conjecturemarton} proving the polynomial Freiman-Ruzsa conjecture
was formalized before the review process  for the final publication had concluded.\footnote{Formalization completed on 5-12-2023: \url{https://mathstodon.xyz/@tao/111526765350663641} 
Referee reports received on 24-4-2024: \url{https://mathstodon.xyz/@tao/112333043706335214}}
In certain branches of theoretical computer science, formalizing work is not only expected, but a natural
part of the research process, because of the substantial mechanical detail needed for proofs.
Thus, these developments suggest that teaching formalization to mathematicians on a larger scale is worthwhile.

We propose a framework for educational formalization projects. An overview of the framework is available in \cref{tab:framework}.
Given the imbalance between capable teachers and potential students, we provide a framework that minimizes teacher input.
This imbalance is not just in terms of number of people, and applies more broadly than the academic setting.
In the Lean Zulip,\footnote{https://leanprover.zulipchat.com} the six most active streams are (with expected number of messages
per week\footnote{retrieved on March 8 2025})
``new members'' (730), ``mathlib4'' (670), ``lean4'' (380), ``general'' (330), ``rss'' (310) and ``Is there code for X?'' (250).
Note that ``new members'' and ``Is there code for X?'' are streams that largely consist of more experienced community members
helping newer ones. This implies that a significant chunk of effort goes towards onboarding newer members in the community and
the process of formalization.
Hence, the imbalance is also a factor in community-driven education, further strengthening the need for a framework
that facilitates learning formalization with minimal teacher input.

This goal is achieved by structuring projects in two distinct phases:
Getting to an initial formalization and getting to a polished formalization.
This structure matches the learning process of the student, because
these goals correspond with lower-order learning goals and higher-order learning goals, respectively.
This means that students are enabled to first grasp the basics before moving on to the more advanced material.
For teachers, the structure clearly indicates how to direct their efforts. 
In the first phase, their role is restricted to providing a good starting point and recommending resources.
The more labor-intensive task of reviewing formalizations is relegated to the second phase.
Postponing this task until the student is ready to take full advantage of the feedback
helps to concentrate teacher effort where it is most effective.
Since teacher input is minimized, this framework offers a way to train a large number of 
formalizers while placing minimal strain on available teachers.

\begin{table}[h!]
  \centering
  \begin{tabular}{m{2.5cm}|m{5cm}|m{5cm}}
  & \textbf{Phase 1} & \textbf{Phase 2} \\
  \hline
  \textbf{Deliverable} & Initial formalization & Polished formalization \\
  \hline
  \textbf{Learning goals} & 
    \begin{itemize}
      \item Working with an ITP
      \item Proving goals
      \item Formulating intermediate goals
    \end{itemize}
   & 
    \begin{itemize}
      \item Refactoring formal proofs
      \item Architecting formal proofs
    \end{itemize}
   \\
  \hline
  \textbf{Teacher role} & 
  \begin{itemize}
    \item Provide goal statement
    \item Recommend resources
  \end{itemize} & 
  \begin{itemize}
    \item Review formalization
  \end{itemize}
  \\
  \hline
  \textbf{Student role} & 
  \begin{itemize}
    \item Focus on learning
  \end{itemize} &
  \begin{itemize}
    \item Attention for details
    \item Attention for structure
    \item Finish product
  \end{itemize}  \\
  \end{tabular}
  \caption{Framework overview}
  \label{tab:framework}
\end{table}

\subsection{Related work}

Formalization of graph theory remains an active research area.
However, until now, Tutte's theorem had not been formalized in Lean.
We provide an overview of recent developments in formalization of graph theory in various systems.
Formalizations of graph theory in Rocq include Dilworth's theorem, Hall's marriage theorem and the Erd\H{o}s-Szekeres theorem 
in 2017 by Singh~\cite{singh2017formalization} and the Weak Perfect Graph Theorem in 2020 by Singh and Natarajan~\cite{singh2020constructive}.
In Isabelle/HOL multiple libraries for graph theory have been published: Graph Theory with a release in 2013 by Noschinkski~\cite{noschinski2015graph},
which contains directed graphs, Undirected Graph Theory with a release in 2022 by Edmonds~\cite{edmonds2022Undirected_Graph_Theory-AFP}.
The latter has a transitive dependency on the former. The latter is also the basis
for the formalization of the Balog–Szemer\'edi–Gowers Theorem by Koutsoukou-Argyraki, Bak{\v{s}}ys and Edmonds~\cite{koutsoukou2023formalisation}.
In 2024 Prieto-Cubides defended his thesis ``Investigations in Graph-theoretical Constructions in Homotopy Type Theory''~\cite{prieto2024investigations},
for which he developed a formalization of graph theory in Agda using univalent foundations.
In Lean, the simple graph library is parth of mathlib. In 2020, Hall's marriage theorem was formalized by Gusakov, Mehta and Miller~\cite{gusakov2021formalizing}.
Of the three presented variants the version for indexed families of sets was merged into mathlib and consequently
ported to Lean 4. 

Most relevant to this work is the formalization of Tutte's theorem~\cite{abdualziz2024isabelleGraphLibrary} which was developed (to our knowledge) in parallel and independently
in Isabelle/HOL by Abdulaziz.
This development is part of an ongoing project focusing on graph algorithms, 
including a proof of Edmonds' Blossom Algorithm~\cite{abdulaziz2024formalcorrectnessproofedmonds}.

Relevant work in the educational context differs in various ways from this work.
Our work focuses on teaching the actual formalization process, whereas most existing works consider
teaching mathematics using formalization.
Thoma and Iannone have studied the effect of learning Lean on the characteristics of students' 
proofs~\cite{thoma2022learning}. The same has been done using Waterproof by Hoofd, Sch\"uler-Meyer
and Wemmenhoven (Chapter 3 of \cite{wemmenhove2025waterproof}). Waterproof is built on top
of Rocq and uses controlled natural language.
The work of Massot on Verbose Lean~\cite{massot:LIPIcs.ITP.2024.27} also uses controlled natural
language. The Mechanics of Proof by Macbeth~\cite{macbeth2024mechanics} offers a good example
of material that is useful prior to doing a formalization project with our framework.
Using proof assistants is also one technique in the broader field of formal methods in software engineering.
Spickova and Zamansky~\cite{spichkova2016teaching} present an
overview of approaches to teach formal methods in that context.
We note that the use of proof assistents for education has a rich history, going back
to Mizar, used to teach logic by Trybulec~\cite{trybulec1983system}.
We refer to Tran Minh, Gonnord and Narboux~\cite{minh2024proof} for a more comprehensive overview.

\subsection{Reading guide}

The structure of this work is as follows: \cref{sec:intro} contains a brief overview of Tutte's theorem and related work.
\cref{sec:mathlib} presents the formalization of Tutte's theorem.
\cref{sec:lessons} contains the framework along with the process of Tutte's theorem as an example.
These two sections can be read relatively independently: \cref{sec:lessons} sometimes refers to
\cref{sec:mathlib} for specific details concerning examples, but these are not necessary to follow
the educational content. 
\cref{sec:future} concludes and suggests ideas for future projects in this area.

\section{Formalization}\label{sec:mathlib}

We present the formalization of Tutte's theorem. First we present a brief overview of prerequisites that were already available
in mathlib before the project. Then we present extensions of mathlib that were part of the formalization. Finally we
present the proof itself.
We focus on the structure and definitions in the formalization and have omitted most proofs using \lean|sorry|,
in addition we have modified code samples for clarity. The names of the results link to the actual complete code in mathlib's GitHub repository.
All code snippets before \cref{subsec:formalization} are from mathlib version 4.1.7. All code snippets after that section
are from a commit on a branch of mathlib: \verb|0d2016d6b2de4c164766a24bce95ca948950844c|. This commit is part of the final pull request to make Tutte's theorem
available in mathlib.

\subsection{Preliminaries}

We provide a brief overview of definitions in mathlib's \verb|SimpleGraph| namespace that are relevant to the proof of Tutte's theorem.
The definitions in this section where already available in mathlib at the start of this project. 
The definitions in \cref{subsec:aug} and onward were added as part of this work.

\subsubsection{SimpleGraph, Subgraph and coercions} \label{subsec:defs}

A simple graph is defined as a symmetric and irreflexive relation on a type \verb|V| (\cref{lst:simplegraph}). 
By default, the \lean|symm| and \lean|loopless| fields are assigned via the tactic \lean|aesop_graph|. 
Aesop is a configurable, tree-based proof search tactic~\cite{limperg2023aesop}. 
For brevity, the definition of \lean|aesop_graph| is omitted; it mostly involves configuring \lean|aesop| with a ruleset tailored to simple graphs. 
Additionally, the intro rule is configured to unfold with default transparency, and the tactic is set to fail if it cannot complete the goal. 
This configuration supports the intended use case: automatically attempting to prove the symmetry and irreflexivity of the given adjacency relation.

\begin{lstlisting}[label={lst:simplegraph}, caption={Definition of SimpleGraph.}]
structure /*!\ml{Basic.lean\#L87}{SimpleGraph}!*/ (V : Type u) where
    Adj : V → V → Prop
    symm : Symmetric Adj := by aesop_graph
    loopless : Irreflexive Adj := by aesop_graph
\end{lstlisting}

Next, we examine the definition of subgraphs.
Subgraphs depend on the graph from which they arise and, 
consequently, also on the type of vertices \verb|V| (\cref{lst:subgraph}).
The \lean|verts| field represents the set of vertices on which the subgraph lives.
This allows specifying whether a particular subgraph is considered with or without certain isolated vertices.
The \verb|adj_sub| field characterizes
the fact that it is a subgraph. The remaining fields just ensure compatibility: 
\verb|edge_vert| enforces that the set of vertices is compatible with the relation given,
and \verb|symm| enforces the symmetry. It is not necessary to include irreflexivity, since this
is derivable from \verb|adj_sub|. Henceforth, 
we will refer to this graph \verb|G| as the ambient graph and the type \verb|V| as 
the ambient vertices in the context of a particular subgraph.

\begin{lstlisting}[label={lst:subgraph}, caption={Definition of Subgraph. }]
structure /*!\ml{Subgraph.lean\#L59}{Subgraph}!*/ {V : Type u} (G : SimpleGraph V) where 
    verts : Set V
    Adj : V → V → Prop
    adj_sub : ∀ {v w : V}, Adj v w -> G.Adj v w
    edge_vert : ∀ {v w : V}, Adj v w -> v ∈ verts
    symm : Symmetric Adj := by aesop_graph
\end{lstlisting}

We now discuss the three basic conversions between graphs and subgraphs (\cref{lst:conversions}):
\lean|coe|, \lean|spanningCoe| and \lean|toSubgraph|.
The two coercions, \verb|coe| and \verb|spanningCoe|, differ only in the vertex type of the resulting \verb|SimpleGraph|. 
These coercions yields a graph on the vertices of the subgraph and
on the ambient vertices, respectively. In the case that the subgraph spans the ambient vertices, the two resulting graphs are equivalent.
Using \verb|toSubgraph|, a \verb|SimpleGraph| can
be interpreted as a \verb|Subgraph| of another. The comparison for \lean|SimpleGraph| originates
from the definition of the distributive lattice structure on the type.
There, for two simple graphs \lean|H| and \lean|G| it is defined that \lean|H ≤ G| is notation for \lean|∀ a b, H.Adj a b → G.Adj a b|.
This condition is the same as the \lean|adj_sub| field in the \lean|Subgraph| structure.

\begin{lstlisting}[label={lst:conversions}, caption={Conversions between graphs and subgraphs. }]
def /*!\ml{Subgraph.lean\#L132}{Subgraph.coe}!*/ (G' : Subgraph G) : SimpleGraph G'.verts where
  Adj v w := G'.Adj v w
  symm _ _ h := G'.symm h
  loopless v h := loopless G v (G'.adj_sub h)

def /*!\ml{Subgraph.lean\#L161}{Subgraph.spanningCoe}!*/ (G' : Subgraph G) : SimpleGraph V where
  Adj := G'.Adj
  symm := G'.symm
  loopless v hv := G.loopless v (G'.adj_sub hv)

def /*!\ml{Subgraph.lean\#L560}{toSubgraph}!*/ (H : SimpleGraph V) (h : H ≤ G) : G.Subgraph where
  verts := Set.univ
  Adj := H.Adj
  adj_sub e := h e
  edge_vert _ := Set.mem_univ _
  symm := H.symm
\end{lstlisting}

\subsubsection{Matchings}

To formulate Tutte's theorem, we first define a perfect matching. All definitions in this section belong to
the \verb|SimpleGraph.Subgraph| namespace. A subgraph satisfies the predicate \verb|IsMatching| 
if for every vertex of the subgraph, there exists
a unique adjacent vertex within that subgraph. A perfect matching is then defined as a subgraph that is
both a matching and a spanning subgraph. 
The support of a subgraph is defined as the domain of the adjacency relation (\cref{lst:matchings}). 
Hence, the sets \verb|verts| and \verb|support| differ precisely by the isolated vertices included in the subgraph.
Because \verb|IsMatching| is based on \verb|verts| instead of \verb|support|, these sets coincide for matchings.
All these definitions are stated in \cref{lst:matchings}.

\begin{lstlisting}[label={lst:matchings}, caption={Definitions of (perfect) matchings and support.}]
def /*!\ml{Matching.lean\#L64}{IsMatching}!*/ (M : Subgraph G) : Prop := 
    ∀ ⦃v⦄, v ∈ M.verts → ∃! w, M.Adj v w

def /*!\ml{Subgraph.lean\#L149}{IsSpanning}!*/ (G' : Subgraph G) : Prop := ∀ v : V, v ∈ G'.verts

def /*!\ml{Matching.lean\#L206}{IsPerfectMatching}!*/ (M : G.Subgraph) : Prop := 
    M.IsMatching ∧ M.IsSpanning

def /*!\ml{Subgraph.lean\#L201}{support}!*/ (H : Subgraph G) : Set V := Rel.dom H.Adj

theorem /*!\ml{Matching.lean\#L208}{IsMatching.support\_eq\_verts}!*/ (h : M.IsMatching) : 
    M.support = M.verts := by sorry
\end{lstlisting}

\subsubsection{Walk, Trail, Path and Cycle}

The definitions of walks, trails, paths and cycles are essential for two main reasons. 
First, they are used to define the notion of reachability, which, in turn, is used to define connected components. 
Second, they play a key role in augmenting matchings, as explained in \cref{subsec:aug}. 
This latter application is the primary motivation for introducing the concepts of walks and cycles in this context.
A \verb|Walk| is defined as an inductive type dependent on the graph \verb|G| in which it lives (\cref{lst:walk}). It strongly 
resembles the standard definition of a list, except that the adjacency condition for consecutive vertices is built into the \verb|cons| constructor. 

\begin{lstlisting}[label={lst:walk}, caption={Definition of walks.}]
inductive /*!\ml{Walk.lean\#L53}{Walk}!*/ (G : SimpleGraph V) : V → V → Type u
  | nil {u : V} : Walk u u
  | cons {u v w : V} (h : G.Adj u v) (p : Walk v w) : Walk u w
  deriving DecidableEq
\end{lstlisting}

When considering a walk, there are several relevant properties for Tutte's theorem. 
To define them, we need two functions: \lean|edges| returns a list of edges as elements of \lean|Sym2 V| and
\lean|support| returns a list of vertices.
The \lean|Nodup| predicate ensures that these lists contain no duplicates.
This is used in the predicates on walks, which are expressed as structures with the relevant properties (\cref{lst:walkhier}). 
\verb|IsTrail| enforces that no edges are duplicated, while \verb|IsPath| enforces that no vertices are duplicated.
\verb|IsCircuit| refers to a nontrivial trail with the same start and end vertices
and \verb|IsCycle| is a circuit with no duplicate vertices apart from the first one.

\begin{lstlisting}[label={lst:walkhier}, caption={Properties of walks.}]
structure /*!\ml{Path.lean\#L78}{IsTrail}!*/ {u v : V} (p : G.Walk u v) : Prop where
  edges_nodup : p.edges.Nodup

structure /*!\ml{Path.lean\#L83}{IsPath}!*/ {u v : V} (p : G.Walk u v) extends IsTrail p : Prop where
  support_nodup : p.support.Nodup

structure /*!\ml{Path.lean\#L91}{IsCircuit}!*/ {u : V} (p : G.Walk u u) extends IsTrail p : Prop where
  ne_nil : p ≠ nil

structure /*!\ml{Path.lean\#L99}{IsCycle}!*/ {u : V} (p : G.Walk u u) extends IsCircuit p : Prop where
  support_nodup : p.support.tail.Nodup
\end{lstlisting}

\subsubsection{Reachability and Connected Components}

Connected components are needed for both the statement and proof of Tutte's theorem. 
First, we define reachability, then we define connected components in terms of reachability (\cref{lst:reach_comp}).
Reachability between vertices \lean|u| and \lean|v| is defined as the type of walks between them
being nonempty, which is a non-constructive way of stating their existence.  This definition allows using and proving the reachability relation by converting
to and from \lean|Walk| respectively. Connected components are then defined as the quotient of the reachability relation.
When dealing with vertices in connected components, it can be helpful to treat the component as a \lean|Set V|, using \verb|supp| (short for support).

\begin{lstlisting}[label={lst:reach_comp}, caption={Reachability and connected components.}]
def /*!\ml{Path.lean\#L804}{Reachable}!*/ (u v : V) : Prop := Nonempty (G.Walk u v)

def /*!\ml{Path.lean\#L975}{ConnectedComponent}!*/ := Quot G.Reachable

def /*!\ml{Path.lean\#L1134}{ConnectedComponent.supp}!*/ (C : G.ConnectedComponent) := 
    {v | G.connectedComponentMk v = C}
\end{lstlisting}

\subsection{Augmenting matchings} \label{subsec:aug}

A common operation on matchings is to extend or 
modify them using the symmetric difference with an alternating path or an alternating cycle, respectively.
In mathlib, both \verb|SimpleGraph| and \verb|Subgraph| have a definition
of the symmetric difference. However, the definition on subgraphs has the quirk that it modifies
the \verb|verts| on which it is defined. In the context of Tutte's theorem, we want to retain all vertices and
only modify the edges, which is precisely what the symmetric difference on graphs is defined to do. This means
that we will be using \verb|SimpleGraph V| as the core type when reasoning about this, despite the fact
that the  \verb|IsMatching| predicate is defined on subgraphs. 
Therefore, we will need to coerce the subgraphs involved to simple graphs. 
As explained in \cref{subsec:defs}, this can be done using either \verb|coe| and \verb|spanningCoe|.
The resulting graphs have different types, since for a subgraph \lean|M| 
it holds that \lean|M.coe : SimpleGraph M.verts| and \lean|M.spanningCoe : SimpleGraph V|.
In the case of perfect matchings, this subgraph is spanning. This means
that \lean|M.verts| is actually \lean|Set.univ|, the set of all ambient vertices.
However, as types, \lean|SimpleGraph M.verts| and \lean|SimpleGraph V| are not definitionally equal, merely equivalent.
To avoid issues with type checking, we use \verb|spanningCoe|, thereby continuing to use \verb|V| as ambient vertices.

In \cref{lst:cyc}, we present the definition of \verb|IsCycles|. While we will only need
to take the symmetric difference with a single cycle, all results apply to sets of cycles.

\begin{lstlisting}[label={lst:cyc}, caption={Definition of IsCycles.}]
def /*!\ml{Matching.lean\#L336}{IsCycles}!*/ (G : SimpleGraph V) := ∀ ⦃v⦄, 
  (G.neighborSet v).Nonempty → (G.neighborSet v).ncard = 2
\end{lstlisting}

\cref{lst:alt} contains the definition of \verb|IsAlternating|. A graph \verb|G| is alternating
with respect to some other graph \verb|G'| if exactly every other edge in \verb|G| belongs to \verb|G'|.
Note that this can only hold if the degree of each vertex in \verb|G| is at most two, because having degree three or more 
forces two incident edges to be either both included or both excluded from \verb|G'|.
This requirement is met by any graph that satisfies \verb|IsCycles|.
The lemma then shows that taking the symmetric difference with a cycle alternating with a perfect matching again yields a perfect matching.

\begin{lstlisting}[label={lst:alt}, caption={Alternating graphs}]
def /*!\ml{Matching.lean\#L527}{IsAlternating}!*/ (G G' : SimpleGraph V) := ∀ ⦃v w w': V⦄, w ≠ w' → 
    G.Adj v w → G.Adj v w' → (G'.Adj v w ↔ ¬ G'.Adj v w')

lemma /*!\ml{Matching.lean\#L562}{IsPerfectMatching.symmDiff\_of\_isAlternating}!*/ (hM : M.IsPerfectMatching)
  (hG' : G'.IsAlternating M.spanningCoe) (hG'cyc : G'.IsCycles) :
  (⊤ : Subgraph (M.spanningCoe ∆ G')).IsPerfectMatching := by sorry
\end{lstlisting}

\subsection{Tutte's theorem formalized} \label{subsec:formalization}

We present the formalization in a top-down manner, 
beginning with the final proof before covering the various components. 
\cref{lst:tutte} presents the statement
and proof of Tutte's theorem. First, we formalize the notion of a Tutte violator and then use that to state Tutte's theorem.
We focus on the main version of Tutte's Theorem, which concerns finite graphs, encoded with
the instance \lean|Fintype V|. We first dismiss the necessity with a lemma, show that
\lean|Fintype.card V| must be even, and then proceed to the core of the proof: showing the
sufficieny of the condition for a perfect matching.

\begin{lstlisting}[label={lst:tutte}, caption={Statement and proof of Tutte}]
def /*!\mlb{Tutte.lean\#L31}{IsTutteViolator}!*/ (G: SimpleGraph V) (u : Set V) : Prop :=
  u.ncard < ((⊤ : G.Subgraph).deleteVerts u).coe.oddComponents.ncard

theorem /*!\mlb{Tutte.lean\#L327}{tutte}!*/ [Fintype V] : (∃ (M : Subgraph G) , M.IsPerfectMatching) ↔ 
    (∀ (u : Set V), ¬ G.IsTutteViolator u) := by
  classical
  refine ⟨by rintro ⟨M, hM⟩; apply not_IsTutteViolator hM, ?_⟩
  contrapose!
  intro h
  by_cases hvOdd : Odd (Fintype.card V)
  · exact ⟨∅, isTutteViolator_empty hvOdd⟩
  · exact exists_TutteViolator h (Nat.not_odd_iff_even.mp hvOdd)
\end{lstlisting}

In \cref{lst:tutte_easy}, we examine the easier parts of the proof. 
In \lean|isTutteViolator_empty| we show that the empty set is a Tutte violator. 
This hinges on \ml{Connectivity/WalkCounting.lean\#L250}{\lean|odd_card_iff_odd_components|}, 
which states that the vertex set has odd cardinality precisely when the graph has an 
odd number of odd components.
In \lean|not_IsTutteViolator| we show the necessity of the condition, by constructing an
injective function from the odd components to the deleted vertices. This is done using
a lemma that shows that under a perfect matching in the original graph, in each odd
component at least one node must be matched to a deleted vertex.
Injectivity follows from the fact that each deleted vertex can only be matched to one other vertex.

\begin{lstlisting}[label={lst:tutte_easy}, caption={The easier parts of the proof}]
theorem /*!\mlb{Tutte.lean\#L139}{isTutteViolator\_empty}!*/ (hodd : Odd (Fintype.card V)) : 
      G.IsTutteViolator ∅ := by sorry

lemma /*!\mlb{Tutte.lean\#L149}{not\_IsTutteViolator}!*/ {M : Subgraph G} (hM : M.IsPerfectMatching) (u : Set V) : ¬G.IsTutteViolator u := by sorry
\end{lstlisting}

In \cref{lst:tutte_suff}, we address the sufficiency. First, we show that it suffices
to consider \lean|Gmax|, an edge-maximal extension of \lean|G|. 
The lemma \ml{Matching.lean\#L322}{\lean|exists_maximal_isMatchingFree|} uses the ordering on graphs, 
the fact that there are only finitely many graphs on a finite vertex set, 
and a general result from mathlib to obtain an edge-maximal graph, and use monotonicity of
the number of odd components to show that it suffices to consider that case.
We then consider two cases: when \lean|G.deleteUniversalVerts| has only cliques as components, 
and when it does not.
For the first case, we quickly defer to a lemma that encapsulates this result.
For the second case, we first obtain certain vertices and properties required to construct
a matching before delegating the details to a lemma.

\begin{lstlisting}[label={lst:tutte_suff}, caption={The sufficiency}]
def /*!\mlb{UniversalVerts.lean\#L32}{universalVerts}!*/ (G : SimpleGraph V) : Set V := 
    {v : V | ∀ ⦃w⦄, v ≠ w → G.Adj w v}

def /*!\mlb{UniversalVerts.lean\#L41}{deleteUniversalVerts}!*/ (G : SimpleGraph V) : Subgraph G := (⊤ : Subgraph G).deleteVerts G.universalVerts

lemma /*!\mlb{Tutte.lean\#L284}{exists\_TutteViolator}!*/ (h : ∀ (M : G.Subgraph), ¬M.IsPerfectMatching)
  (hvEven : Even (Fintype.card V)) :
  ∃ u, G.IsTutteViolator u := by
\end{lstlisting}

In \cref{lst:tutte_cliques}, we address the case in which the graph decomposes into cliques. 
First, we construct a matching that covers 
all components by matching one vertex from each odd component to a universal vertex.
Then, we show that if we remove the matched nodes, each component remains with an even number of vertices.
This allows a matching where all remaining vertices within each component are matched internally.
In \lean|exists_of_isClique_supp| it is established that the remaining vertices are even in number and to 
obtain a matching on those vertices. These to matchings are then joined to yield a perfect matching.

\begin{lstlisting}[label={lst:tutte_cliques}, caption={The case of cliques}]
theorem /*!\mlb{Tutte.lean\#L122}{Subgraph.IsPerfectMatching.exists\_of\_isClique\_supp}!*/
  (hveven : Even (Fintype.card V)) (h : ¬G.IsTutteViolator G.universalVerts)
  (h' : ∀ (K : G.deleteUniversalVerts.coe.ConnectedComponent),
  G.deleteUniversalVerts.coe.IsClique K.supp) : ∃ (M : Subgraph G), M.IsPerfectMatching := by
  sorry
\end{lstlisting}

If the graph does not decompose into cliques, we first obtain two near matchings (as matchings on a graph with one added edge).
In \cref{lst:tutte_not_cliques}, we use the symmetric difference of these matchings to find an 
alternating cycle primarily contained in this symmetric difference (and fully within \lean|G|). 
This is done by obtaining an alternating path within symmetric difference, starting with the edge \lean|s(a, c)|. 
If \lean|x| cannot be reached, we obtain an alternating cycle for immediate use; otherwise, the path ends at either \lean|x| or \lean|b|.
Both these cases are then dismissed using a helper lemma.

Note that this theorem has a relatively large number of hypotheses.
Some of these hypotheses are essential and would also appear in an informal proof as arguments, 
whereas the rest merely encode assumptions about adjacency and vertex distinctness. 
In this context, we primarily work with \lean|V| as ambient vertices and use subgraphs of \lean|G| wherever feasible.
As noted in \cref{subsec:aug}, the symmetric difference
for \lean|Subgraph| is not suitable in this context. 
Consequently, whenever possible, we use \lean|spanningCoe| to convert a subgraph to a \lean|SimpleGraph G|.

\begin{lstlisting}[label={lst:tutte_not_cliques}, caption={The case of non-cliques}]
  private theorem /*!\mlb{Tutte.lean\#L165}{tutte\_exists\_isPerfectMatching\_of\_near\_matchings}!*/ {x a b c : V}
  {M1 : Subgraph (G ⊔ edge x b)} {M2 : Subgraph (G ⊔ edge a c)} (hxa : G.Adj x a)
  (hab : G.Adj a b) (hnGxb : ¬G.Adj x b) (hnGac : ¬ G.Adj a c) (hnxb : x ≠ b) (hnxc : x ≠ c)
  (hnac : a ≠ c) (hnbc : b ≠ c) (hM1 : M1.IsPerfectMatching) (hM2 : M2.IsPerfectMatching) :
  ∃ (M : Subgraph G), M.IsPerfectMatching := by sorry
\end{lstlisting}

Since we have now treated all the cases, this completes the formalization.

\section{Framework for educational formalization projects}\label{sec:lessons}

This section presents our framework for educational formalization projects. We first describe our framework
and embed it in Bloom's taxonomy to link it to the theory of learning. This embedding shows 
that the framework supports training beginners to undertake an extensive formalization project. We then propose conditions
for the successful application of the framework. 
Throughout, we explain how this framework minimizes teacher input.
Finally, we illustrate the framework by applying it to the process of formalizing Tutte's theorem. 
The overview of the framework appears
in \cref{tab:framework} (in \cref{sec:intro}).

The goal of the framework is to facilitate the process of learning formalization while 
minimizing teacher input. The student's learning is prioritized, the product of the project is secondary.
However, a capable student could produce a formalization that benefits the broader community, within the scope of the project.

\subsection{Framework description}

The framework consists of two phases. In the first phase, the student produces an initial formalization, 
which is then polished (and optionally integrated) in the second phase.
These phases align with lower-order and higher-order thinking skills in the revised Bloom's taxonomy~\cite{anderson2001taxonomy}. With this framework we aim
for teaching procedural knowledge related to formalization. Working with an interactive theorem prover, proving goals
and formulating goals are part of understanding and applying the interactive theorem prover, corresponding to the second and third
category in Bloom's taxonomy. 
Hence, the first phase mainly fosters lower-order thinking skills. In order to
successfully refactor and architect formal proofs, a student needs the higher-order thinking skills
of analysis, the fourth category in Bloom's taxonomy. A good student will also grow to self-assess
their refactored proof, for which they need to be able to evaluate the quality. This is part of the fifth category of Bloom's taxonomy.
Thus, our framework spans the second through fifth categories of Bloom's taxonomy, facilitating a student's progression from 
beginner to executing more extensive formalization projects.

The teacher role adapts throughout the project. 
During the first phase, the primary task is to provide the student with a correct goal statement with an appropriate scope.
It is crucial that the details 
of the goal statement are accurate; the goal should be provable and correspond to the mathematical idea
that is communicated to the student. When the student encounters difficulties in the initial phase, 
focus on providing tools. Preferably, this involves referencing a resource that 
enables them to resolve the problem independently. If relevant resources are lacking, focus on explaining how you would overcome the difficulty 
and guide them through that process, rather than providing the answer. This prevents repetitive questions
and thus minimizes teacher input in the long run. During the second phase, the teacher role shifts to being a reviewer. 
This provides the student with examples of important details and proper structuring of the formal proof. 
In this phase, the teacher will
offer more specific pointers and guide the student with targeted advice, rather than teaching a general workflow.

The teacher should also communicate the expectations regarding the student role. It is recommended to emphasize that
the initial goal is to get a working formalization. Students should be informed not to worry excessively about issues 
concerning the structure of the proof, duplicated code, and general quality. Note that this does not imply a complete
disregard for these issues. However, if structural or quality issues prevent the student from completing the
proof, then these issues should be addressed.
A practice that can be recommended is leaving TODO markers when the student signals a quality issue that is not
problematic for getting to an initial formalization.
Students should be made aware that addressing these issues is the goal of the second phase.
It is important for students to understand that they can spend a relatively large amount of time thinking 
about how something should be structured, compared to the time needed to implement it.
The scope of the projects should be limited enough to enable capable students to produce a 
formalization adhering to the standards of the respective community.

This framework is both compatible with communities with 
centralized and decentralized development models. In case of a more centralized 
development model, such as mathlib or Rocq's mathcomp, a capable student will get to 
submit pull requests to the centralized repository. In case of a more decentralized model,
such as the Archive of Formal Proofs for Isabelle, the second phase might involve
splitting part of the proof into a reusable library and submitting these 
separately to the AFP.

\subsection{Conjectured necessary conditions} \label{subsec:conditions}

Since we present only one example, we cannot be certain of the exact conditions to successfully
apply this framework. We propose some necessary conditions that we conjecture to be important.

For the student, we propose one condition: The student should be at the appropriate level.
This means the student should have a basic understanding of formalization. 
For a student with some mathematical experience, that could be achieved by working
through an extensive tutorial such as `Theorem Proving in Lean' \cite{avigad2021theorem}.
For a less experienced student, teaching the mathematics and formalization in parallel is an option,
for example, using `The Mechanics of Proof'~\cite{macbeth2024mechanics}. 
In addition, we suggest that the student should not have mastered the skills taught in the first phase.
Once a student has acquired these skills, it becomes much more feasible to aim for a polished formalization 
immediately. This condition ensures that the student's level matches the learning goals in our framework.

The teacher must be able to fulfill three responsibilities: pointing to relevant learning resources, 
reviewing the proof, and selecting a suitable goal. 
Pointing to the relevant resources in the first phase is important for
minimizing teacher input. If the teacher cannot defer to resources,
they must instead spend time explaining the issue at hand.
Being able to review the proof is essential for the learning process. This aspect of the framework
helps students eventually tackle more substantial projects independently.
Selecting a suitable goal consists of two parts: selecting a mathematical idea to be formalized
and formalizing the statement. The latter part is relatively straightforward; the teacher should
formalize the goal mainly in terms of existing definitions. New definitions may be introduced, but the student should be informed that 
they are free to modify them if it benefits the formalization, as long as the mathematical content remains unchanged.
Selecting a suitable idea depends on the context. In a traditional academic setting, 
the most important part is that the formalization can be completed in the allotted time.
We suggest a rule of thumb: If the initial formalization is can be completed in half the allotted time, this will leave enough
time for the second phase. It also provides some flexibility and should ensure every student has something to be assessed on, 
even if it is not a fully polished formalization. This rule of thumb is based on the timeline of the formalization of
Tutte's theorem (see \cref{subsec:tutte_ex}).
In a community-driven context, the teacher should confer with the student about what they consider a suitable goal.
In general, a suitable target for formalization is something that is valuable to the community, 
is not currently being pursued by others and should not lie on the critical path for larger efforts.
This arrangement allows the student to progress at their own pace while contributing something valuable to the community.

\subsection{Tutte's theorem as an educational formalization project}\label{subsec:tutte_ex}

We describe the formalization of Tutte's theorem as an application of our
proposed framework. This project is an instance of independent community-driven education:
the author began the formalization as a learning endeavor, with a potential contribution to mathlib.
The Lean community fulfilled the teacher role, primarily through asynchronous communication
through the Lean Zulip\footnote{https://leanprover.zulipchat.com} and in GitHub pull request comments.

We show how the proposed conditions emerged from the execution of this particular project.
Prior to this project, the author's experience with formalization was limited to tutorials, a formalization
of the multinomial theorem in Lean 3, and the porting of several files from Lean 3 to Lean 4 in mathlib.
This satisfies the requirement for a basic understanding of formalization, without implying extensive experience.
The mathematical idea to formalize was selected by the author, based on a TODO marker in mathlib indicating
it was a desired result. Kyle Miller helped to arrive at a correct formal statement to target for the first
phase. Ya\"el Dillies took on a very large part of the teacher role in the second phase by consistently reviewing
the pull requests in the combinatorics area, prior to mathlib maintainers doing the final review.
In these final reviews, Bhavik Mehta and other maintainers provided a lot of useful feedback.
The ability to point to relevant resources was partially fulfilled.
The community would refer to important parts of ``Theorem Proving in Lean''~\cite{avigad2021theorem}.
However, some resources did not yet exist and therefore could not be referenced.
The Lean Language Reference~\cite{leanlanguageref} is one such resource that did not exist at the start of
the project and would have been helpful at an earlier stage.
Since this was an instance of independent education, the scope was not tied to any particular timeline or
workload. The first phase was started in September of 2023 and finished on 16 July 2024.
The second phase then ran from July 2024 to March 2025. Note that due to varying time commitments
during this time, we cannot state that the lead time of the phases is proportional to the effort.
We estimate that the time was split equally over the two phases, which led us to the rule of thumb presented in \cref{subsec:conditions}:
aim for the first phase to take half the allotted time.

Since the formalization is nearly fully integrated in mathlib, the author 
self-assesses that the learning goals outlined in the framework have been achieved. 
Working with an ITP, proving goals and formulating intermediate goals have both been clearly
demonstrated. At the end of the first phase, the formalization was contained in a single file with 5757 lines. 
At the time of writing, the formalization
contains 686 lines, spread over 2 files. The remainder of the formalization was contributed to mathlib or 
superseded by improved proofs, reducing its overall size. We claim that this demonstrates the ability to 
refactor and architect formal proofs. Documenting all insights in detail would be overly verbose for this work.
Instead, we illustrate the type of insights students should gain from a project in our framework, using two examples.

\subsubsection{Example: have-tactic pattern}\label{subsec:have-tac}

The lessons in the first phase will be relatively basic. One example is the use
of the have-tactic pattern, where intermediate goals are stated, and a 
tactic leveraging local hypotheses is then used to discharge the goal.
Although this pattern is not always the best way to present a polished proof,
we consider it to be a useful tool in achieving an initial formalization.
It encapsulates the idea that the user of the ITP provides motivation for
the reason why something is true, followed by some tactic that automates
the ``trivial'' part of the proof. \cref{lst:havetactic} contains an example
that uses the \lean|omega| tactic for linear arithmetic, but this pattern
is also useful in conjunction with more specific tactics (like \lean|exact| or \lean|apply|)
or more general-purpose tactics (such as \lean|aesop|).
We remark that this pattern is an example of something that is well-supported by
the Isabelle/Isar framework by Wenzel~\cite{wenzel2002isabelle}. While the Lean syntax has the ability
to support Isar-style proofs, it is not enforced or broadly recommended. 
We hypothesize that a similar framework could be developed for Lean and
this might help students to structure their proofs in a more readable way.

\begin{lstlisting}[label={lst:havetactic}, caption={have-tactic pattern}]
have : (Fintype.card V + 1) - (p.length + 1 + 1) < (Fintype.card V + 1) - (p.length + 1) := by
  have h1 := SimpleGraph.Walk.IsPath.length_lt hpp
  omega
\end{lstlisting}

\subsubsection{Example: Abstraction of representatives}

The second phase allows more in-depth lessons, given that it concerns higher-order learning skills.
We present one such example along with the broader learnings we draw from it.
We describe an architectural decision made during integration into mathlib and discuss the associated trade-offs.
In the proof shown in \cref{lst:tutte_cliques}, we aim to obtain exactly one vertex from 
each odd component that remains after deleting universal vertices.
Concretely, we first take the set of odd components, then consider the image under \lean|Quot.out| followed by \lean|Subtype.val|.
\lean|Quot.out| produces a vertex in the connected component. \lean|Subtype.val|
converts this vertex from \lean|G.deleteUniversalVets.coe| to \lean|V|. 
The key property of this construction is that it provides exactly one vertex from each odd component, 
with no vertices outside those components.
We present four versions of formalization of this property,
with various levels of abstraction: 
Version 1 abstracts the underlying set of components. This
version does not admit other choices for representatives, which makes it less reusable. Version 2 abstracts the choice of representatives.
Version 3 refines Version 2 by consolidating the properties into a single statement, leveraging
the bijectivity of the function that maps a vertex to its corresponding component (a strategy suggested by Eric Wieser). 
Version 4 then abstracts to the setting
of \lean|Quot| rather than \lean|ConnectedComponent|.

\begin{lstlisting}[label={lst:rep}, caption={Versions of representatives of connected components. Version 3 from mathlib, others inspired from older versions.}]
-- Concrete set of representatives
def oddVerts (G : SimpleGraph V) : Set V := Subtype.val '' (Quot.out '' 
    G.deleteUniversalVerts.coe.oddComponents)
-- Version 1
lemma rep_unique {C : Set (G.ConnectedComponent)} (c : G.ConnectedComponent) 
    (h : c ∈ C) : ∃! v, v ∈ Quot.out '' C ∩ c.supp := by sorry

lemma disjoint_rep_image_supp {C : Set (G.ConnectedComponent)} (c : G.ConnectedComponent) 
    (h : c ∉ C) : Disjoint (Quot.out '' C) c.supp := by sorry
-- Version 2
Set.Represents (s : Set V) (C : Set G.ConnectedComponent) where
  unique_rep {c : G.ConnectedComponent} (h : c ∈ C) : ∃! v, v ∈ s ∩ c.supp
  exact_rep {c : G.ConnectedComponent} (h : c ∉ C) : s ∩ c.supp = ∅

lemma represents_of_image_exists_rep_choose (C : Set G.ConnectedComponent) :
    ((fun c ↦ c.out) '' C).Represents C where
  unique_rep {c} h := by sorry
  exact_rep {c} {h} := by sorry
-- Version 3
def /*!\ml{Represents.lean\#L28}{Represents}!*/ (s : Set V) (C : Set G.ConnectedComponent) : Prop :=
  Set.BijOn G.connectedComponentMk s C

lemma /*!\ml{Represents.lean\#L35}{image\_out}!*/ (C : Set G.ConnectedComponent) :
    Represents (Quot.out '' C) C := by sorry
-- Version 4
def Represents (s : Set α) (C : Set (Quot r)) := Set.BijOn (Quot.mk r) s C

lemma out_image_represents (C : Set (Quot r)) : (Quot.out '' C).Represents C := by sorry
\end{lstlisting}

If we take a broader perspective, we observe three factors at play. Firstly, there is abstraction with respect to the set of representatives.
Secondly, we have abstraction along the type axis: \verb|Quot| versus \lean|ConnectedComponent|. Finally, there is abstraction along a mathematical axis,
thinking in terms of functions rather than vertices. Our assessment of whether these abstractions are worthwhile depends on different
criteria in each case. For the first factor, the main concern is reusability. We can conceive of situations where the choice of 
representatives does matter. Abstracting the set of representatives does not have a material impact on the proof in this case.
This makes it a cheap abstraction. For deciding on abstraction along the
type axis, in addition to reusability, wider library-related concerns played a role. In the end, the generic version was
rejected based on concerns of misuse in other contexts, partially due to the accompanying \lean|SetLike| instance.
The abstraction along the mathematical axis is practically a free lunch: its clean formulation results in a quick proof
on the default set of representatives. The cost is that the element-wise properties need separate proofs, but these are only needed for the abstract definition. 

We hypothesize that an abstraction along the type axis is still worthwhile. However, including it would require identifying the
circumstances in which this notion is more useful than misleading. In generic software engineering, the ``rule of three'' is often applied. 
This rule, popularized by Martin Fowler~\cite{fowler2018refactoring}, states that refactoring code for abstraction needs three instances of use:
The first usage is a concrete implementation. In the second usage, the relevant code is copied and modified. 
In the third usage, the common parts of the three usages are then abstracted away and consolidated to a single piece of code. 
We propose that this rule is also appropriate for this case. The fact that the definitions can be converted to \lean|Quot r| wholesale
shows that abstraction is possible. The remaining question is: What are the other use cases? Depending on where they are, this
would inspire the formulation and location of the abstract version. For example, if the other uses are all in the combinatorics
part of the library, then that also would where to place the abstraction. If other areas of the library adopt this approach,
then placing the abstraction somewhere in the \lean|Data| namespace might be more appropriate.
Given that mathlib currently has 1.7 million lines of code, detecting similar patterns that could be abstracted seems a non-trivial task.
Search engines could help with this. In the case of strict type-based search, such as Loogle\footnote{\url{https://loogle.lean-lang.org/}}, 
providing an option to show results both more and less specific than the types given could help identify these patterns.
Alternatively, more fuzzy approaches, like LeanSearch~\cite{gao2024semantic}, seem very appropriate here: The similarity based on the mathematical ideas
could allow the identification of the similar patterns amenable to abstraction.
We propose that these improvements would facilitate easier refactoring efforts to arrive at the best abstraction.

\section{Conclusion and future work}\label{sec:future}

In conclusion, we presented the formalization of Tutte's theorem, a key theorem in matching theory. This formalization
has largely been contributed to mathlib, providing a stepping stone towards the formalization of research-level mathematics
in this area. We presented a framework for educational formalization projects, applicable in both traditional academic and independent community-driven settings.
With this framework, we offer teachers a way to teach more advanced formalization efficiently, by minimizing the input required from them.

Some interesting smaller projects to extend the treatment of simple graphs in mathlib include adding the graph formulation of Hall's marriage theorem 
by building on Gusakov et al.~\cite{gusakov2021formalizing} or proving the Tutte--Berge formula. An more ambitious project might be inspired by
the ongoing work of Abdulaziz~\cite{abdulaziz2024formalcorrectnessproofedmonds} and prove
various results for graph algorithms in Lean. 
Regarding education, an interesting project would apply the presented framework on a larger scale and use the resulting
feedback to propose improvements. This could be done both in the traditional academic and independent community-driven settings.
A Isabelle/Isar-style framework for Lean could also be developed, as suggested in \cref{subsec:have-tac}.

\bibliography{main}

\end{document}